\documentclass[aps,pre,reprint,groupedaddress]{revtex4-1}

\usepackage{graphicx}
\usepackage{subfig}

\usepackage{color}

\usepackage{bm}
\usepackage{amsmath}
\usepackage{amsfonts}
\usepackage{amssymb}
\usepackage[latin1]{inputenc}

\newcommand{\ket}[1]{|#1\rangle}

\def\Tr{{\rm Tr}}
\def\sys{{\rm sys}}

\begin{document}

\title{Analytical expression for variance of homogeneous-position quantum walk with decoherent position }
\author{Mostafa Annabestani}
\email{annabestani@shahroodut.ac.ir}
\affiliation{Department of Physics, Shahrood university of technology, Shahrood, Iran}

\begin{abstract}
We have derived an analytical expression for variance of homogeneous-position decoherent quantum walk (HPDQW) with general form of noise on its position, and have shown that, while the quadratic ($t^2$) term of variance never changes by position decoherency, the linear term ($t$) does and always increases the variance. We study the walker with ability to tunnel out to $d$ nearest neighbors as an example and compare our result with former studies. We also show that, although our expression have been derived for asymptotic case, the rapid decay of time-dependent terms cause the expressions to be correct with a good accuracy even after dozens of steps.   
\end{abstract}




\maketitle

\section{Introduction}
\label{sec:intro}
Quantum walks are actually the quantum  counterparts of the classical random walks. The main difference between these two, comes from very fundamental quantum properties such as superposition and interference \cite{Nayak}.
Two types of quantum walks have been introduced: discrete time quantum walks (DTQW) and continuous time quantum walks (CTQW) \cite{Kempe03,Venegas-Andraca2012}. The relation between these two was not clear for a long time, but is known nowadays \cite{Childs2010}. 

Quantum walks have attracted much attention in the recent years, not only because they are exploited to design powerful quantum algorithms \cite{Childs04,Shenvi03,Hillery2012}, but also they can be used as a universal model for quantum computations \cite{Childs2009,UniversalUnderwood2010}.
Quantum walks have been investigated from different points of views such as their propagation speed \cite{Kempe02,Kendon2003,Konno2}, entanglement between coin and position of the walker \cite{annabestani,Abal06-ent,Romanelli2014}, quantum walks in higher dimensions \cite{Mackay} and quantum walks as an entanglement generator \cite{Venegas-entanglementgeneration}.

Even though different experiments are conducted for quantum walks \cite{Knight-OC,Ryan,Zhao-prop}, more advanced and applied experiments are practically challenging, one of the most important problem is the interaction of quantum walk system with its environment.
These interactions can eliminate the quantum properties of the system and change it to a classical one, therefore many research topics are  focused  on decoherency and the effects of noise on quantum walks \cite{Kendon-review,Brun03,lopez,zhang,annabestani-deco}. Especially it is proven that noise in the coin space eliminates the quantum properties of the walk and changes it to a  classical random walk\cite{Brun2,Brun03}. 
Although weak noise can enhance the properties of quantum walk in such a way that to be useful and desirable for developing quantum algorithms \cite{KT03}, it is shown that increasing the noise level, eliminates all quantum properties and changes it to a classical system \cite{Brun2}.
One may concludes that any kind of noise can eliminate the quantum properties and change the system to a classical one, but it is shown that if the environmental noise allows the walker to tunnel out into its closest neighbor with a defined probability, not only the quantum walk does not transit to a classical one, but also it helps to improve quantum properties \cite{Annabestani2016}.

In this paper we assume general form of noise on the position of  one dimensional homogeneous-position quantum walk and investigate whether the system makes transition from quantum to classic. We show that for any kinds of noise on the position of walker, the system keeps its quantum properties. We also find an exact analytical form of variance in the presence of general noise on position as well as a numerical calculation of coin-position entanglement (CPE) as two witnesses of our claim.

This work is organized as follows. Sec.\ref{background} gives a
brief review on the  structure of one-dimensional QW and Kraus representation for study of decoherency. We have devoted Sec.\ref{moments} to introduce the analytical expressions for the first and second moments in the presence of general noise. By assuming general form of noise on the position of a quantum walker, an analytical compact formula for variance of one-dimensional quantum walk with general initial state is derived  in Sec.\ref{positionDecoherency}. This formula  is then used in Sec.\ref{example} to analyze a quantum walker with noise on its position which enables it to tunnel out to its neighbors. We summarize our results and present our conclusions in Sec.\ref{conclusion}.

\section{Background}\label{background}
Quantum Walk (QW) on a line is defined as a quantum system with two Hilbert spaces, the coin space $H_c$ spanned by $\{|L\rangle,|R\rangle\}$, and the position space  $H_p$, spanned by $\{|i\rangle \mid  i=-\infty,\cdots,\infty \}$. For each steps of walking, we need to apply the following unitary operator
\begin{equation}\label{U-w}
U_w=S\;(I\otimes U_c),
\end{equation}
where $U_c \in U(2)$ is the coin operator and
\begin{equation}\label{S}
S=\sum\limits_x {\left| {x + 1} \right\rangle \left\langle x \right|
\otimes \left| R \right\rangle \left\langle R \right| + \left| {x -
1} \right\rangle \left\langle x \right| \otimes \left| L
\right\rangle \left\langle L \right|},
\end{equation}
is conditional shifting operator. Therefore the evolution of system is 
\begin{equation}\label{t-step-waking}
|\Psi\left(t+1\right)\rangle={U_w}|\Psi\left(t\right)\rangle
\,\,\rightarrow
\,\,|\Psi\left(t\right)\rangle={U_w}^t|\Psi\left(0\right)\rangle.
\end{equation}

This equation, in fact, describes the coherent evolution of the system, but in
practice, it is impossible to isolate the system from its
environment. So, unavoidable interaction with the environment affects the coherent evolution of the system and decoherency occurs.

One of the most important approaches which is useful to investigate the decoherency, 
is the Kraus representation \cite{Nielsen}. In order to review it briefly, we define $H_E$ as the Hilbert space of the environment, spanned by $\{\ket{e_n} \mid n = 0\cdots m\}$ where $m+1$ is the dimension of it. The system is not isolated from its environment, so the whole system (system + environment) is defined on a larger Hilbert space ($H=H_W\otimes H_E$) which evolving together. One who only needs the state of the system, can obtain it by tracing over the environment's degrees of freedom
\begin{equation}\label{rho-sys}
    \rho_{sys}=\Tr_{env} \left( U \rho\, U^\dagger \right).
\end{equation}
Without loss of generality, we assume that the states of the system ($\rho_0$) and the environment ($|\chi_0\rangle\langle \chi_0|$) are initially unentangled ($\rho=\rho_0\otimes|\chi_0\rangle\langle \chi_0|$). So, we
can write Eq. (\ref{rho-sys}) as
\begin{equation}\label{KR-rho-rep-1}
\rho_{\sys}=\sum\limits_{n = 0}^m {\left\langle {e_n }
\right|U\left| {\chi_0 } \right\rangle \rho _0 \left\langle {\chi_0
} \right|U^\dag \left| {e_n } \right\rangle }  = \sum\limits_{n =
0}^m {E_n \rho_0 E_n^\dag  },
\end{equation}
where $\left\{E_n=\left\langle {e_n} \right|U\left| {\chi_0 }
\right\rangle \mid n=0,1,\cdots,m \right\}$ are Kraus operators. These operators satisfy the following completeness relation \cite{Nielsen}
\begin{equation}\label{compeletness-relation}
\sum\limits_n{{E_n}^\dag E_n }=I.
\end{equation}

By definition of Kraus operators, one step of walking can be
written as 
\begin{equation}\label{first-step-rho}
\rho\left(t+1\right)=\sum\limits_{n = 0}^m {E_n \rho\left(t\right)
E_n^\dagger}, 
\end{equation}
and the state of the walker after $t$ steps is
\begin{equation}\label{KR-rho-rep-t}
\rho\left(t\right)=\sum\limits_{n_t = 0}^m {...\sum\limits_{n_2 =
0}^m {\sum\limits_{n_1 = 0}^m
{E_{n_t}...E_{n_2}E_{n_1}\rho\left(0\right) E_{n_1}^\dagger
E_{n_2}^\dagger ...E_{n_t}^\dagger}}}.
\end{equation}

This equation is general and one can obtain Kraus operators $E_n$ and use this equation to find the final state of the walker. Unfortunately this equation is more complicated than that can be used for analytical calculations, but in some cases it can be simplified using some mathematical tricks. Brun \cite{Brun03} has been derived a compact analytical formula for the first and the second moments of probability distribution of one dimensional quantum walk in the presence of coin only decoherence and Annabestani et al \cite{annabestani-deco} have generalized Brun approach and have found a compact formula for moments in general form of decoherence. In the next section we briefly explain their approach.\\

\section{Moments of decoherent quantum walk}\label{moments}
$E_n$  are operators that act on the system (coin+position) Hilbert space. Therefore one can write the general form of $E_n$ as follows

\begin{equation}\label{general-En}
\begin{split}
E_n  &= \sum\limits_{x,x'}{\sum\limits_{i,j} {a_{x,x',i,j}^{(n)}
\left| {x'} \right\rangle \left\langle x \right| \otimes \left| i
\right\rangle \left\langle j \right|} }\\ &= \sum\limits_x
{\sum\limits_l {\sum\limits_{i,j} {a_{x,l,i,j}^{(n)} \left| {x + l}
\right\rangle \left\langle x \right| \otimes \left| i \right\rangle
\left\langle j \right|} } },
\end{split}
\end{equation}
where $x,l=-\infty,\cdots,\infty$ and $i,j=\{L,R\}$.

In \cite{annabestani-deco}, we have shown that if the coefficients $a^{(n)}_{x,l,i,j}$ do not depend on $x$, ($a^{(n)}_{x,l,i,j}\equiv a^{(n)}_{l,i,j}$) (this is why we call it homogeneous-position QW ) then we can derive analytical expressions for the first and second moments of
position as follows
\begin{widetext}
\begin{align}\label{moments-1,2-final-form}
\begin{split}
{\langle x\rangle} _t &=  i \int_{ - \pi }^\pi \frac{dk}{{2\pi}}
\sum\limits_{m = 1}^t \Tr\left\{ {{\cal G}_k \left( {{\cal L}_k^{m -
1} \left| {\psi _0 \rangle \langle \psi _0 } \right|} \right)}
\right\} \\ 
\left\langle {x^2 } \right\rangle_t  &=
\int_{ - \pi }^\pi  \frac{dk}{2\pi}\sum\limits_{m = 1}^t
\sum\limits_{m' = 1}^{m - 1} \Tr\Big\{ \mathcal{G}_k^\dag
\mathcal{L}_k^{m - m' - 1} \left( \mathcal{G}_k \mathcal{L}_k^{m' -
1} \left|\psi_0\rangle\langle\psi_0\right|  \right)+
\mathcal{G}_k \mathcal{L}_k^{m - m' - 1} \left( \mathcal{G}_k^\dag
\mathcal{L}_k^{m' - 1} \left|\psi_0\rangle\langle\psi_0\right|
\right)\Big\}  \\
 &+ \int_{ - \pi }^\pi
{\frac{dk}{{2\pi}}\sum\limits_{m = 1}^t {\Tr\left\{
{\mathcal{J}_{k}\left( {\mathcal{L}_{k}^{m - 1}
\left|\psi_0\rangle\langle\psi_0\right| } \right)}  \right\}}}.
\end{split}
\end{align}
\end{widetext}
in this equation
\begin{align}\label{L-k}
\mathcal{L}_{k} \tilde{O}&=\left.\mathcal{L}_{k,k'} \tilde{O}\right|_{k'=k}\\\label{G-k}
\mathcal{G}_{k} \tilde{O}&=\left.\mathcal{G}_{k,k'} \tilde{O}\right|_{k'=k}
\end{align}
\begin{equation}\label{J}
\mathcal{J}_{k}\tilde{O}=\left.\frac{{d\mathcal{G}_{k,k'}^\dag\tilde{O}
}}{{dk}}\right|_{k'=k}=\left.\sum\limits_n {\frac{{dC_n \left( k
			\right)}}{{dk}}\tilde{O}\frac{{dC^{\dag}_n \left( k'
			\right)}}{{dk'}}} \right|_{k'=k}
\end{equation}
where
\begin{eqnarray}\label{L-k,k'}
  \mathcal{L}_{k,k'} \tilde{O}&=& \sum\limits_n{C_n\left(k\right) {\tilde{O}} C_n^\dag\left(k'\right)}\\\label{G}
\mathcal{G}_{k,k'}\tilde{O}&=&\sum\limits_n {\frac{{dC_n \left( k
\right)}}{{dk}}\tilde{O}C^{\dag}_n \left(k'\right)}
\end{eqnarray}
with
\begin{equation}\label{Cn}
C_n \left( k \right) = \sum\limits_l {\sum\limits_{i,j} {a_{l,i,j}^{(n)} e^{ - ilk} \left| i \right\rangle \left\langle j \right|}. }
\end{equation}

In \eqref{moments-1,2-final-form} the initial state has been assumed as (in $k$-space)
\begin{equation} \rho _0  = \iint {\frac{{dkdk'}}{{4\pi ^2 }}\left|
k \right\rangle \left\langle {k'} \right| \otimes \left| {\psi _0 }
\right\rangle \left\langle {\psi _0 } \right|}.
\end{equation}
\section{Noise on  position of HPDQW}\label{positionDecoherency}
In this section, we will show that if only the position of HPDQW is subject to decoherence, its variance never transits to classical one. This is the main result of our paper and we will derive general form of variance in the presence of position decoherency as well. 

Let us define our model as a case in which only the position of walker is affected by the environmental interactions. So one step of walking can be written as   
\begin{equation}\label{first-step-rho-positionOnly}
\rho\left(t+1\right)=\sum\limits_{n = 0}^m {P_n U \rho\left(t\right)
U^\dagger P_n^\dagger},
\end{equation}
where $U$ is responsible for the coherent evolution of walker defined in \eqref{U-w} and $P_n$ is a general operator on the position subspace, which can be written as
\begin{align}\label{general-Pn}
P_n  = \sum\limits_{x^\prime,x}
{{ {p_{x,x^\prime}^{(n)} \left| {x^\prime }
\right\rangle \left\langle x \right| } } }\otimes I=\sum\limits_{x,l}
{{ {p_{x,l}^{(n)} \left| {x + l}
\right\rangle \left\langle x \right| } } }\otimes I.
\end{align}

We restrict our attention to the one dimensional quantum walk with homogeneous position ($p_{x,l}^n \equiv p_{l}^n$) as Annabestani et al assumed in \cite{annabestani-deco}, so
\begin{eqnarray}\label{general-Pn-homogenous}
P_n  = \sum\limits_{x,l}
{{ {p_{l}^{(n)} \left| {x + l}
\right\rangle \left\langle x \right| } } }  \otimes I.
\end{eqnarray}
By using Fourier transformation
\begin{equation} \label{Four-x}
\left| x \right\rangle  = \int\limits_{ - \pi }^\pi
{\frac{{dk}}{{2\pi }}e^{ - ikx} \left| k \right\rangle },
\end{equation}
$P_n$ has a diagonal form in $k$-space as
\begin{eqnarray}\label{general-Furier-Pn-homogenous}
\tilde{P}_n  = \int\limits_{ - \pi }^\pi
{{ \frac{{dk}}{{2\pi }}F_n\left(k\right)\left| {k}
			\right\rangle \left\langle k \right| } }  \otimes I,
\end{eqnarray}
where 
\begin{eqnarray}\label{Fn}
F_n(k)  = \sum\limits_{l}
 {p_{l}^{(n)}e^{-ilk}}.
\end{eqnarray}

From \eqref{first-step-rho-positionOnly} it is clear that
\begin{equation}\label{P_nU}
E_n=P_nU.
\end{equation}
So by using the explicit form of $U$ in \eqref{U-w},\eqref{S} and $P_n$ in \eqref{general-Pn-homogenous} it is not difficult to calculate
\begin{eqnarray}\label{Cn-kspace}
C_n(k)=F_n(k)U(k)
\end{eqnarray}
where $F_n(k)$ is defined in \eqref{Fn} and
\begin{eqnarray}
U\left( k \right) = \left( {\begin{array}{*{20}{c}}
{{e^{ - ik}}}&0\\
0&{{e^{ik}}}
\end{array}} \right){U_C}.
\end{eqnarray}

We should note that $E_n$ in \eqref{P_nU} is non diagonal in $x$-space, while it is block-diagonal in $k$-space with blocks $C_n(k)$. So completeness relation of \eqref{compeletness-relation} implies that
\begin{equation}\label{compeletness-relation-C_n}
\sum\limits_n{{C_n(k)}^\dag C_n(k) }=I,
\end{equation}
which it means
\begin{equation}\label{compeletness-relation-F_n}
\sum\limits_n {F_n^* F_n }=\sum\limits_n {F_n F_n^*}=1.
\end{equation}

In order to finding $\mathcal{G}_{k,k'}$ and $\mathcal{J}_{k}$ in \eqref{G},\eqref{J} we need
\begin{equation}\label{d-Ck}
\frac{{d{C_n}(k)}}{{dk}} = {\dot F_n}\left( k \right)U\left( k \right) - i{F_n}\left( k \right)ZU\left( k \right)
\end{equation}
where ${\dot F_n}\left( k \right)={{dF\left( k \right)} \mathord{\left/
 {\vphantom {{dF\left( k \right)} {dk}}} \right.
 \kern-\nulldelimiterspace} {dk}}$ and $Z$ is Pauli matrix $\sigma_z$.
By inserting this equation and \eqref{Cn-kspace} into \eqref{L-k}-\eqref{J}, we have
\begin{align}
\begin{split}
\mathcal{L}_k &= \sum\limits_n {F_n U \hat O F_n^* U^\dag}=\left(\sum\limits_n {F_n F_n^*}\right) U \hat O U^\dag\\ 
\mathcal{G}_k &= \sum\limits_n {\left( {{{\dot F}_n}U - i{F_n}ZU} \right)\hat O\left( {{F_n}^*{U^\dag }} \right)} \\
&= \sum\limits_n {\left( {F_n^*{{\dot F}_n} - iZ_L} \right)U} \hat O{U^\dag }\\
\mathcal{J}_k &= \sum\limits_n {\left( {{{\dot F}_n}U - i{F_n}ZU} \right)\hat O\left( {\dot F_n^*{U^\dag } + iF_n^*{U^\dag }Z} \right)} \\
 &= \sum\limits_n {\left( {\dot F_n^*{{\dot F}_n} + iF_n^*{{\dot F}_n}{Z_R} - i\dot F_n^*{F_n}{Z_L} + {Z_L}{Z_R}} \right)U} \hat O{U^\dag }
\end{split}
\end{align}
where $Z_L\hat O\equiv Z\hat O$ and $Z_R \hat O \equiv \hat O Z$.

For simplicity we define $\left | F\right \rangle\equiv \vec V$ as a complex vector with elements $V_i=F_i(k),\,\,\, i=0\dots m$ and $\mathcal{W}_k \hat O\equiv U(k)\hat O U^{\dag}(k)$. By these definitions 
\begin{align}\label{superOperator-simplified}
\begin{split}
\mathcal{L}_k &= \mathcal{W}_k\\
\mathcal{G}_k &= \left( {\langle {F}|\dot F\rangle - i{Z_L}} \right)\mathcal{W}_k\\
\mathcal{J}_k &= \left( \langle \dot{F}| \dot{F}\rangle +2Re\left(i\langle {F}|\dot F\rangle {Z_R}\right) + {Z_L}{Z_R} \right){\mathcal{W}_k}
\end{split}
\end{align}
in which we use \eqref{compeletness-relation-F_n}.

In this paper we use Bloch representation \cite{Nielsen}, in which any two-by-two density matrix can be represented by four-dimensional column vector $\vec{r}$ as
\begin{equation}
\hat{O}=\sum_{i=0}^{3}{r_i \sigma_i} \equiv \left( {\begin{array}{*{20}c}
	{r_0 }  \\
	{r_1 }  \\
	{r_2 }  \\
	{r_3 }  \\
	\end{array}} \right),
\end{equation}
where $\sigma_i$ are Pauli matrices and $r_i=Tr(\hat{O}\sigma_i)/2$.
In this representation any trace preserve quantum operator $\epsilon$ is equivalent to a map of the form
\begin{equation}
\epsilon(\hat{O})=\hat{O^\prime} \equiv \vec{r}\rightarrow \vec{r^\prime}=M\vec{r}+\vec{c},
\end{equation}
where $M$ is a four-by-four matrix and $\vec{c}$ is a constant vector.

Using Bloch representation it is easy to see that
\begin{equation}
{Z_R} = Z_L^* \equiv \left( {\begin{array}{*{20}{c}}
0&0&0&1\\
0&0&{i}&0\\
0&-i&0&0\\
1&0&0&0
\end{array}} \right)
\end{equation}
and trace preserve super operator $\mathcal{W}_k$ is 

\begin{equation}\label{W_k}
\mathcal{W}_k\equiv\left( \begin {array}{*{20}c} 1&0&0&0\\
0&0& \sin \left(2k \right) & \cos \left( 2k \right)\\
0&0& -\cos \left( 2k \right)& \sin \left( 2k\right) \\
0&1&0&0  \end {array} \right) 
\end{equation}

By inserting Super operators of \eqref{superOperator-simplified} into \eqref{moments-1,2-final-form} and assuming general form of initial state as
\begin{equation}
\left| {{\psi _0}} \right\rangle \left\langle {{\psi _0}} \right| = \left( {\begin{array}{*{20}{c}}
{1/2}\\
{{r_1}}\\
{{r_2}}\\
{{r_3}}
\end{array}} \right)
\end{equation}

\begin{align}
\begin{split}
{\langle x\rangle} _t &=  i \int_{ - \pi }^\pi \frac{dk}{2\pi}
\sum\limits_{m = 1}^t \Tr\left\{{ {\left( {\langle {F}|\dot F\rangle - i{Z_L}} \right)} {{\cal W}_k^{m} \left| {\psi _0 \rangle \langle \psi _0 } \right|}}\right\}\\
&= 2i \left( {\begin{array}{*{20}c}
   {\langle {F}|\dot F\rangle} & {0} & {0} & {-i}  \\
\end{array}} \right)\Gamma_1\left(t\right)
\left( {\begin{array}{*{20}{c}}
{1/2}\\
{{r_1}}\\
{{r_2}}\\
{{r_3}}
\end{array}} \right) 
\end{split}
\end{align}
where 
\begin{equation}\label{Gamma1}
\Gamma_1\left(t\right) =  \int_{ - \pi }^\pi \frac{dk}{2\pi}
\sum\limits_{m = 1}^t {\cal W}_k^{m}.
\end{equation}

 In order to find the second moments we use these facts that, $\langle F | \dot F \rangle$ is pure imaginary (see \eqref{Fn}) and $\mathcal{G}_k^\dag=\mathcal{G}_k^*$. So the first row of $\mathcal{G}_k$ and $\mathcal{G}_k^\dag$ are equal with different sign which can be useful to simplify the first term of $\langle x^2 \rangle_t$ in \eqref{moments-1,2-final-form} as
\begin{widetext}
\begin{equation}\label{Integral1}
\begin{split}
I_1 &= 2\int_{ - \pi }^\pi  \frac{dk}{2\pi}\sum\limits_{m = 1}^t
\sum\limits_{m' = 1}^{m - 1} \left( {\begin{array}{*{20}{c}}
{\langle{F}|\dot F\rangle}&0&0&-i
\end{array}} \right)
\mathcal{W}_k^{m - m'} \left( \mathcal{G}_k^\dag-\mathcal{G}_k  \right)\mathcal{W}_k^{m' -
1} \left|\psi_0\rangle\langle\psi_0\right|\\
&= 2\int_{ - \pi }^\pi  {\frac{{dk}}{{2\pi }}\sum\limits_{m = 1}^t {\sum\limits_{m' = 1}^{m - 1} {\left( {\begin{array}{*{20}{c}}
{\langle{F}|\dot F\rangle}&0&0&-i
\end{array}} \right)\mathcal{W}_k^{m - m'}\left(-2{\langle{F}|\dot F\rangle}+i(Z_L+Z_R) \right)} } } \mathcal{W}_k^{m'}\left( {\begin{array}{*{20}{c}}
{1/2}\\
{{r_1}}\\
{{r_2}}\\
{{r_3}}
\end{array}} \right)\\
&=2{\left( {\begin{array}{*{20}{c}}
{\langle{F}|\dot F\rangle}&0&0&-i
\end{array}} \right)}\int_{ - \pi }^\pi  \frac{{dk}}{{2\pi }}\Big\{-2{\langle{F}|\dot F\rangle} \sum\limits_{m = 1}^t {\sum\limits_{m' = 1}^{m - 1}\mathcal{W}_k^{m}  }+i\sum\limits_{m = 1}^t {\sum\limits_{m' = 1}^{m - 1}\mathcal{W}_k^{m-m'}(Z_L+Z_R)\mathcal{W}_k^{m'}  }\Big\}\left( {\begin{array}{*{20}{c}}
{1/2}\\
{{r_1}}\\
{{r_2}}\\
{{r_3}}
\end{array}} \right).
\end{split}
\end{equation}
\end{widetext}

The calculation of the first term in this integral is straightforward and we just need
\begin{equation}\label{Gamma2}
\Gamma_2(t)=\int_{ - \pi }^\pi  \frac{{dk}}{{2\pi }}\sum\limits_{m = 1}^t {\sum\limits_{m' = 1}^{m - 1}\mathcal{W}_k^{m}  },
\end{equation}
but the second term needs more attention. In order to calculate it, we note that
\begin{equation}
{Z_L} + {Z_R} = \left( {\begin{array}{*{20}{c}}
0&0&0&2\\
0&0&0&0\\
0&0&0&0\\
2&0&0&0
\end{array}} \right)
\end{equation}
so
\begin{align}\label{lastTerm}
\begin{split}
&\sum\limits_{m = 1}^t {\sum\limits_{m' = 1}^{m - 1}\mathcal{W}_k^{m-m'}(Z_L+Z_R)\mathcal{W}_k^{m'}  }\left( {\begin{array}{*{20}{c}}
{1/2}\\
{{r_1}}\\
{{r_2}}\\
{{r_3}}
\end{array}} \right)\\
=&\sum\limits_{m = 1}^t {\sum\limits_{m' = 1}^{m - 1}\mathcal{W}_k^{m-m'}}\left( {\begin{array}{*{20}{c}}
{f(m',k,\vec r)}\\
{{0}}\\
{{0}}\\
{{1}}
\end{array}} \right)\\
=&\sum\limits_{m = 1}^t {\sum\limits_{m' = 1}^{m - 1}\left({\begin{array}{*{20}{c}}
{{f(m',k,\vec r)}}\\
{{0}}\\
{{0}}\\
{{0}}
\end{array}}\right)}+\sum\limits_{m = 1}^t {\sum\limits_{m' = 1}^{m - 1}\mathcal{W}_k^{m-m'}}\left( {\begin{array}{*{20}{c}}
{0}\\
{{0}}\\
{{0}}\\
{{1}}
\end{array}} \right)
\end{split}
\end{align}
where
\begin{equation}
f(m',k,\vec r)=2\left\{\mathcal{W}_k^{m'} \left( {\begin{array}{*{20}{c}}
{1/2}\\
{{r_1}}\\
{{r_2}}\\
{{r_3}}
\end{array}} \right)\right\}_4.
\end{equation}
Note that we can write
\begin{equation}
\sum\limits_{m = 1}^t {\sum\limits_{m' = 1}^{m - 1}\mathcal{W}_k^{m-m'}}\equiv \sum\limits_{m = 1}^t {\sum\limits_{m' = 1}^{m - 1}\mathcal{W}_k^{m'}}
\end{equation}
because when $m'$ goes from 1 to $m-1$,  the power of $m-m'$ changes from $m-1$ to 1, so we can easily reverse the order of terms in the inner summation. So the simple final form of $I_1$ from \eqref{Integral1} can be written as
\begin{equation}
I_1=2{\left( {\begin{array}{*{20}{c}}
{\langle{F}|\dot F\rangle}&0&0&-i
\end{array}} \right)}\vec{\mathcal{V}}
\end{equation}
with
\begin{equation}
\vec{\mathcal{V}}=-2{\langle{F}|\dot F\rangle}\Gamma_2(t)\left( {\begin{array}{*{20}{c}}
{\frac{1}{2}}\\
{{r_1}}\\
{{r_2}}\\
{{r_3}}
\end{array}} \right) +\left( {\begin{array}{*{20}{c}}
{i\gamma(t)}\\
{{0}}\\
{{0}}\\
{{0}}
\end{array}} \right)+\Gamma_2'(t)\left( {\begin{array}{*{20}{c}}
{0}\\
{{0}}\\
{{0}}\\
{{i}}
\end{array}} \right)
\end{equation}
where
\begin{eqnarray}\label{Gamma2p}
\gamma(t)&=&\int_{ - \pi }^\pi  \frac{{dk}}{{2\pi }}\sum\limits_{m = 1}^t {\sum\limits_{m' = 1}^{m - 1}f(m',k,\vec r)  }\\\nonumber
\Gamma_2'(t)&=&\int_{ - \pi }^\pi  \frac{{dk}}{{2\pi }}\sum\limits_{m = 1}^t {\sum\limits_{m' = 1}^{m - 1}\mathcal{W}_k^{m'}  }
\end{eqnarray}
and the last term of $\langle x^2 \rangle_t$ in \eqref{moments-1,2-final-form} will be
\begin{eqnarray}\nonumber
I_2 &=& 2 \left( {\begin{array}{*{20}c}
   {\langle {\dot F}|\dot F\rangle}+1 & {0} & {0} & {2Re(i{\langle {F}|\dot F\rangle})}  \\
\end{array}} \right)\Gamma_1\left(t\right)
\left( {\begin{array}{*{20}{c}}
{1/2}\\
{{r_1}}\\
{{r_2}}\\
{{r_3}}
\end{array}} \right)\\\nonumber
\end{eqnarray}

Now our problem reduces to calculations of $\Gamma_1,\Gamma_2,\Gamma^\prime_2$ and $\gamma$. For this purpose we use spectral decomposition which needs eigenvalues and the corresponding eigenvectors of  $\mathcal{W}_k$. By straightforward calculations, we have 
 \begin{equation}\label{eigenvalues-position}
 \lambda_1=\lambda_2=1 \quad,\quad\lambda_3=e^{i\left(\theta+\pi\right)}\quad,\quad
 \lambda_4=e^{-i\left(\theta+\pi\right)},
 \end{equation}
 and  \small
 \begin{align}\label{eigenvectors-position}
 \left|e_1\right\rangle &=\frac{1}{N_1}\left( \begin {array}{c} \sin \left( k \right) \\\notag
 \cos \left( k \right) \\\sin \left( k \right)
 \\\cos \left( k \right) \end {array} \right) \\
 \left|e_2\right\rangle &=\frac{1}{N_2}\left( \begin {array}{c} -2\left( \cos^2\left(k\right)+1 \right) \\
 \sin \left( 2k \right) \\2\sin^2 \left( k \right)
 \\\sin \left( 2k \right) \end {array} \right) \\
 \left|e_3\right\rangle = \left|e_4\right\rangle^*&=\frac{1}{N_3}\left( \begin {array}{c} 0 \\
 \sin \left( 2k \right)e^{2i\theta} \\1-\cos \left( 2k \right)e^{2i\theta}
 \\-\sin \left( 2k \right)e^{i\theta} \end {array} \right),\notag
 \end{align}
where $\cos\left(\theta\right)=\cos^2\left(k\right)$ and $N_1, N_2, N_3$ are normalization factors. So we can write $\Gamma_1\left(t\right)$ in \eqref{Gamma1} as
\begin{align}\label{Gamma1AB}
\Gamma_1\left(t\right)&=\int^\pi_{-\pi}{\frac{{dk}}{{2\pi }}\sum\limits^t_{m=1}{\mathcal{W}^{m}_k}}\\\notag
&=\int^\pi_{-\pi}\frac{{dk}}{{2\pi }}\sum\limits^t_{m=1}
\Big\{\left|e_1\right\rangle\left\langle
e_1\right|+\left|e_2\right\rangle\left\langle e_2\right|\\\nonumber
&+e^{im\left(\theta+\pi\right)}\left|e_3\right\rangle\left\langle
e_3\right|+e^{-im\left(\theta+\pi\right)}\left|e_4\right\rangle\left\langle
e_4\right|\Big\}\\\nonumber
&=tA-B
\end{align}
where
\begin{align}\label{A}
A&=\int^\pi_{-\pi}{\frac{{dk}}{{2\pi }}\left(\left|e_1\right\rangle\left\langle e_1\right|+\left|e_2\right\rangle\left\langle e_2\right|\right)}\\\notag
&=\left( \begin {array}{*{20}c} 1&0&0&0\\
0& \alpha & 0 &  \alpha \\
0& 0 & 1-2\alpha&0 \\
0&\alpha & 0 & \alpha  \end {array} \right)
\end{align}
with $\alpha=1-\frac{\sqrt{2}}{2}$ and
\begin{align}\label{B}
B&=\int^\pi_{-\pi}\frac{{dk}}{{2\pi }}{2Re{\left(\frac{e^{i\theta}}{1+e^{i\theta}}\left|e_3\right\rangle\left\langle e_3\right|\right)}}\\\notag
&=\left( \begin {array}{*{20}c} 0&0&0&0\\
0& \frac{\sqrt{2}}{4} & 0 &  \frac{3\sqrt{2}}{4}-1 \\
0& 0 & \alpha&0 \\
0&-\frac{\sqrt{2}}{4} & 0 & \frac{\sqrt{2}}{4} \end {array} \right).
\end{align}

%
 For  $\Gamma_2$ in \eqref{Gamma2}, we can write
\begin{align}\label{Gamma2first}\notag
\Gamma_2\left(t\right)&=\int^\pi_{-\pi}{\frac{{dk}}{{2\pi }}\sum\limits_{m = 1}^t {\sum\limits_{m' = 1}^{m - 1}\mathcal{W}_k^{m}}}=\int^\pi_{-\pi}{\frac{{dk}}{{2\pi }}\sum\limits_{m = 1}^t {\left(m-1\right)\mathcal{W}_k^{m}}}\\
&=\int^\pi_{-\pi}{\frac{{dk}}{{2\pi }}\sum\limits_{m = 1}^t {m\mathcal{W}_k^{m}}}-\Gamma_1\left(t\right)
\end{align}
where we have used \eqref{Gamma1AB} for the last term. Calculation of the first term needs more attention and can be written as
\begin{align}\label{sum-mW}
\nonumber
\int^\pi_{-\pi}{\frac{{dk}}{{2\pi}}\sum\limits^t_{m=1}{m\mathcal{W}^{m}_k}}=&\int^\pi_{-\pi}\frac{{dk}}{{2\pi}}\sum\limits^t_{m=1}
m\Big\{\left|e_1\right\rangle\left\langle
e_1\right|+\left|e_2\right\rangle\left\langle e_2\right|\\\nonumber
&+e^{im\left(\theta+\pi\right)}\left|e_3\right\rangle\left\langle
e_3\right|+e^{-im\left(\theta+\pi\right)}\left|e_4\right\rangle\left\langle
e_4\right|\Big\}\\
=&\frac{t}{2}\left(t+1\right)A-C
\end{align}
where
\begin{align}
C&=\int^\pi_{-\pi}{\frac{{dk}}{{2\pi}}2Re{\left(\frac{e^{i\theta}}{\left(1+e^{i\theta}\right)^2}\left|e_3\right\rangle\left\langle e_3\right|\right)}}\\\notag
&=\left( \begin {array}{*{20}c} 0&0&0&0\\
0& \frac{3\sqrt{2}}{16} & 0 &  -\frac{\sqrt{2}}{16} \\
0& 0 & \frac{\sqrt{2}}{8} &0 \\
0&-\frac{\sqrt{2}}{16} & 0 &\frac{3\sqrt{2}}{16} \end {array} \right)
\end{align}
Note that we use the fact that $me^{im\left(\theta+\pi\right)}\equiv -i\frac{\partial}{\partial \theta}e^{im\left(\theta+\pi\right)}$. By putting \eqref{sum-mW} and \eqref{Gamma1AB} into \eqref{Gamma2first} we see that,
\begin{align}
\Gamma_2\left(t\right)=\frac{t}{2}\left(t-1\right)A+B-C.
\end{align}
$\Gamma_2^\prime\left(t\right)$ is the last matrix we need to calculate, 
\begin{align}\label{Gamma2pfirst}\notag
\Gamma_2'\left(t\right)=&\int^\pi_{-\pi}{\frac{{dk}}{{2\pi }}\sum\limits_{m = 1}^t {\sum\limits_{m' = 1}^{m - 1}\mathcal{W}_k^{m'}}}\\
=&\int^\pi_{-\pi}\frac{{dk}}{{2\pi}}\Big\{\frac{t}{2}\left(t-1\right)\left(\left|e_1\right\rangle\left\langle
e_1\right|+\left|e_2\right\rangle\left\langle e_2\right|\right)\\\nonumber
&+2\Re{\left(\frac{-te^{i\theta}}{1+e^{i\theta}}+\frac{e^{i\theta}}{\left(1+e^{i\theta}\right)^2}\left|e_3\right\rangle\left\langle e_3\right|\right)}\Big\}\\
=&\frac{t}{2}\left(t-1\right)A-tB+C
\end{align}
Now by putting everything together
\begin{align}
\sigma^2=\left\langle x^2 \right\rangle-\left\langle x \right\rangle^2=at^2+bt+c,
\end{align}
where
\begin{equation}\label{varianceCoefficients}
\begin{split}
a&=\alpha-4\alpha^{2} \left( r_3+r_1 \right) ^2 \\
b&= 2\sqrt{2}\alpha \left(r_3^2-r_1^2 \right) +{\langle{\dot F}|\dot F\rangle}+{\langle{F}|\dot F\rangle}^2 \\
c&= -\frac{1}{2}\left( r_3-r_1 \right) ^2+\frac{3\sqrt {2}}{8}.
\end{split}
\end{equation}

We see that, the quadratic term of variance (a) does not depend on $\left| F\right\rangle$ which means that, the position decoherency keeps the variance quadratic. On the other hand $\left| F\right\rangle$ appears in the linear term (b) in the form of $\langle \dot{F}|\dot{F} \rangle +\langle F|\dot{F} \rangle^2$ which is positive and increases the linear term $b$. We should note that, although we derive our formula in asymptotic case ($t\rightarrow \infty$), but very fast decay of time-dependent terms implies that \eqref{varianceCoefficients} to be a good approximation even for finite number of steps.
\section{Example}\label{example}
Suppose a quantum walker on a line which is able to move to its right and left neighbors with probabilities $p$ and $q$ respectively. If we assume that the walker only moves to the nearest neighbors, then one step of walking can be  written as
\begin{align}\notag\label{nearestNeighbor}
\rho_{t+1}=&\left(1-\left(p+q\right)\right)U \rho_{t}
	U^\dagger\\
	&+pS_+U \rho_{t}U^\dagger S_+^\dagger+qS_-U \rho_{t} U^\dagger S_-^\dagger
\end{align}
where $S_\pm=\sum_x \left|x\pm 1\right\rangle\left\langle x\right|$ are shifting operators. If the walker is able to move to the next nearest neighbors too, then
\begin{align}\notag\label{nextNearestNeighbor}
\rho_{t+1}=&\left(1-\left(p^2+q^2+2pq\right)\right)U \rho_{t}
U^\dagger\\
&+p^2\mathcal{S}_+^2U \rho_{t} U^\dagger+q^2\mathcal{S}_-^2U \rho_{t} U^\dagger\\\notag
 &+2pq\mathcal{S}_+\mathcal{S}_-U \rho_{t} U^\dagger
\end{align} 
in which we have used super operators $\mathcal{S}_\pm\tilde{O}=S_\pm \tilde{O} S_\pm ^\dagger $ for simplicity. This model can be extended to general case in which the walker can move to $d^{th}$ neighbors. So
\begin{equation}\label{DthNearestNeighbor}
\begin{split}
\rho_{t+1}=&\left(1-\sum_{j=0}^d \left( {\begin{array}{*{20}{c}}
		d\\
		j
		\end{array}} \right)p^{d-j}q^j\right)U \rho_{t} U^\dagger\\
&+\sum_{j=0}^d \left( {\begin{array}{*{20}{c}}
	d\\
	j
	\end{array}} \right)\left(p\mathcal{S}_+\right)^{d-j}\left(q\mathcal{S}_-\right)^j U \rho_{t} U^\dagger.
\end{split}
\end{equation}
 
It is not difficult to show that $\mathcal{S}_+^{d-j}\mathcal{S}_-^j\equiv \mathcal{S}_+^{d-2j}$. If we make some simplifications and back to the operator notations
\begin{equation}\label{DthFinalForm}
\begin{split}
\rho_{t+1}=&\left(1-\left(p+q\right)^d\right)U \rho_{t} U^\dagger\\
&+\sum_{j=0}^d \left( {\begin{array}{*{20}{c}}
	d\\
	j
	\end{array}} \right) p^{d-j}q^j S_+\left(d-2j\right) U \rho_{t} U^\dagger S_+\left(d-2j\right)^\dagger
\end{split}
\end{equation}  
in which we have used $S_+\left(r\right)=\sum_x \left|x+ r\right\rangle\left\langle x\right|$. By comparing \eqref{DthFinalForm} with \eqref{first-step-rho-positionOnly}, \eqref{general-Pn} we have
\begin{equation}
\begin{split}
P_0=&\sqrt{1-\left(p+q\right)^d}\\
P_{j+1}=&\left(\sqrt{\left( {\begin{array}{*{20}{c}}
	d\\
	j
	\end{array}} \right) p^{d-j}q^j}\right) S_+\left(d-2j\right),\quad j=0\dots d
\end{split}
\end{equation}
So from \eqref{general-Pn-homogenous} and \eqref{Fn}
\begin{align}\label{Fi}
\begin{split}
F_0=&\sqrt{1-\left(p+q\right)^d}\\
F_{j+1}=&\left(\sqrt{\left( {\begin{array}{*{20}{c}}
		d\\
		j
		\end{array}} \right) p^{d-j}q^j}\right) e^{-i\left(d-2j\right)k},\quad j=0\dots d
\end{split}
\end{align}
therefore
\begin{align}
\left|F\right\rangle=\left( {\begin{array}{*{20}{c}}
	\sqrt{1-\left(p+q\right)^d}\\
	p^{\frac{d}{2}}e^{-idk}\\
	\vdots\\
	\left(\sqrt{\left( {\begin{array}{*{20}{c}}
			d\\
			j
			\end{array}} \right) p^{d-j}q^j}\right) e^{-i\left(d-2j\right)k}\\
	\vdots\\
	q^{\frac{d}{2}}e^{idk}
	\end{array}} \right)
\end{align}
and finally
\begin{align}
{\langle{F}|\dot F\rangle}=&-i\sum_{j=0}^d{{\left(d-2j\right)\left( {\begin{array}{*{20}{c}}
			d\\
			j
			\end{array}} \right) p^{d-j}q^j}}\\
{\langle{\dot F}|\dot F\rangle}=&\sum_{j=0}^d{{\left(d-2j\right)^2\left( {\begin{array}{*{20}{c}}
			d\\
			j
			\end{array}} \right) p^{d-j}q^j}}
\end{align}
after some calculations we have
\begin{align}
{\langle{F}|\dot F\rangle}=&-id\left(p+q\right)^{d-1}\left(p-q\right)\\
{\langle{\dot F}|\dot F\rangle}=&d\left(p+q\right)^{d-2}\left(d\left(p-q\right)^2+4pq\right)
\end{align}
so the effect of decoherency which appears only in the coefficient $b$ in \eqref{varianceCoefficients} can be written as
\begin{align}\label{G_alpha}
G\left(\beta,d,P\right)&={\langle{\dot F}|\dot F\rangle}+{\langle{F}|\dot F\rangle}^2\\\notag
&=dP^d\left(d\left(2\beta-1\right)^2\left(1-P^d\right)+4\beta\left(1-\beta\right)\right)
\end{align}
in which we rewrite the equation in terms of $\beta$ and $P$ using $p=\beta P$ and $q=\left(1-\beta\right)P$. $P$ is the probability of movement  and $\beta$ is the probability of moving to right in each step.

\begin{figure}
	\centering
	\includegraphics[width=7 cm]{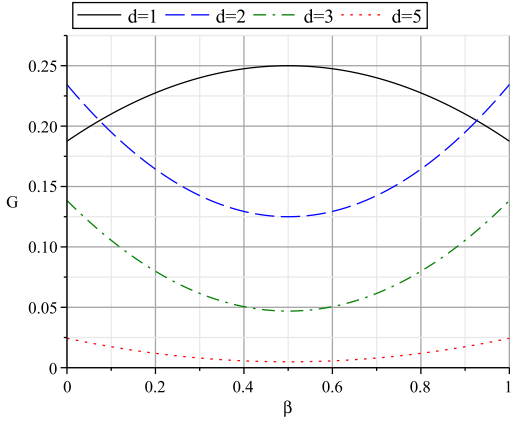}\hfill
	\caption{(Color online) G versus $\beta$ for
		various $d$ and $P=\frac{1}{4}$}.\label{GvsAlphap025}
\end{figure}

\begin{figure}
	\centering
	\includegraphics[width=7 cm]{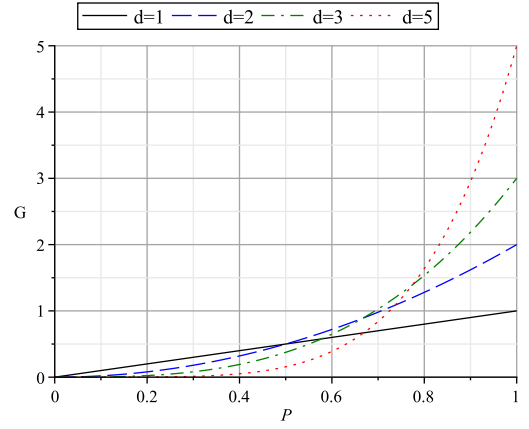}\hfill
	\caption{(Color online) G versus P for
		various $d$ and $\beta=\frac{1}{2}$}.\label{GvsPalphap12}
\end{figure}

Clearly $G\left(\beta,d,P\right)\geq 0$, so any neighbor-movement decoherency increase the variance of probability distribution. $G\left(\beta,d,P\right)$ has an extremum at $\beta=\frac{1}{2}$ (unbiased movement) and it is symmetric around this point (Fig. \ref{GvsAlphap025}).
At this point, $G\left(\beta,d,P\right)$ has very simple form
\begin{align}\label{G_alpha12}
	G\left(\frac{1}{2},d,P\right)=dP^d
\end{align}
this function shows that, for small value of $P$, grater value of $G\left(\frac{1}{2},d,P\right)$ is achievable by smaller $d$ whereas for larger value of $P$, larger $d$ is needed (Fig. \ref{GvsPalphap12}).
 Note that we define $P$ as the probability of one movement . So in the case with $d$ allowable neighbors, we will have $d$-movements terms with probability proportional to $P^d$. From \eqref{DthFinalForm} we can see that the walker with probability $1-P^d$ does not move while it moves and spreads through the $d$ nearest neighbors with probability $P^d$. We define $P_t=P^d$ as the total probability of movements. With $P_t$ we are able to investigate the influence of $d$ in QW. In other words, if the walker with probability $P_t$ moves to neighbors, what the differences are, if it moves farther?
\section{Quantum behavior and speed Up}
The variance and the standard deviation of probability distribution are related to the speed of spreading which is potentially related to the speed of quantum walk based algorithms.
 
How much speed up we have, if the walker is able to move to its neighbors? If we believe that our results are correct only for asymptotic regime, then the answer of this question would be ``nothing" because we can neglect $b$ in $t\to \infty$, but we can show that with excellent accuracy, all results are correct, even for dozens of steps.

\begin{figure}
	\centering
	\includegraphics[width=7.8 cm]{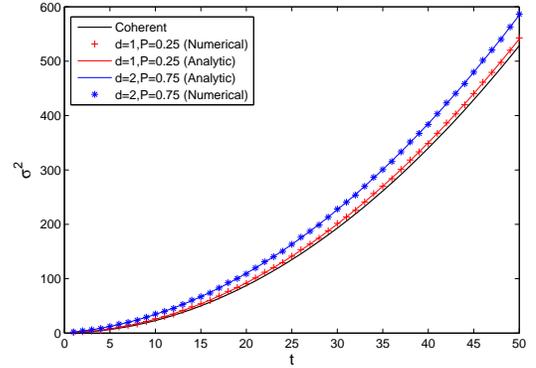}
	\caption{(Color online) $\sigma^2$ versus $t$ for coherent case (lower curve), decoherent case with $d=1$ ,$P=0.25$ (middle curve) and decoherent case with $d=2$ ,$P=0.75$ (upper curve). The discrete symbols are respective numerical results}.\label{NumVsTheor}
\end{figure}

As the evidences for our claim, we simulate some cases and compare them with results \eqref{varianceCoefficients}, see Fig. \ref{NumVsTheor}.
Our simulation shows that theoretical prediction  matches with numerical results within the accuracy of $\pm 0.5\%$ even for 20 steps. So not only neighbor-movement decoherency keeps the quadratic terms of variance same as the coherent QW, but also increases the linear term by $tG$ which may be valuable for finite number of steps. For example in the cases plotted in Fig.\ref{NumVsTheor}, the variance of decoherent walk for $d=2 , P=0.75$ is $24\% (18\%)$ grater than the variance of coherent walk after 20 (30) steps and it is $5\% (3.5\%)$ greater for $d=1 , P=0.25$.

\begin{figure}
	\centering
	\includegraphics[width=7 cm]{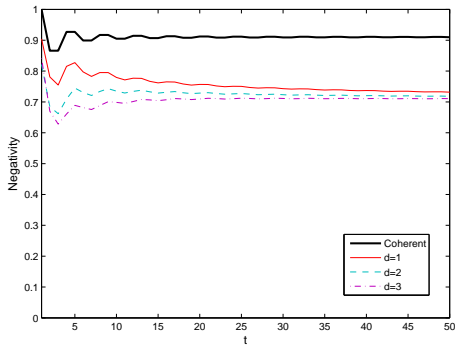}\hfill
	\caption{(Color online) Negativity versus $t$ for $P_t=0.25$ and
		 $d=1,2,3$. Initial state for all cases is $\left|\psi_0\right\rangle=\frac{\left|0\right\rangle+i\left|1\right\rangle}{\sqrt{2}}$ and upper bold curve is negativity of coherent one}.\label{NegVsT}
		
\end{figure}

\begin{figure}
	\centering
	\includegraphics[width=7.8 cm]{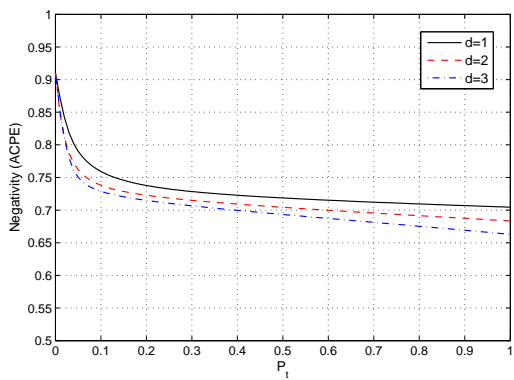}
	\caption{(Color online) Negativity versus $P_t$ after $t=50$
		for $d=1,2,3$. Initial state for all cases is $\left|\psi_0\right\rangle=\frac{\left|0\right\rangle+i\left|1\right\rangle}{\sqrt{2}}$}.\label{NegVsPtot}
\end{figure}

Another important result, which we would like to emphasize, is the saturation of  asymptotic coin-position entanglement (ACPE) in the presence of this decoherency (Fig.\ref{NegVsT}). Our simulations show that increasing $d$, decreases the ACPE but never vanishes it, unlike the coin decoherency \cite{Brun03}, (Fig.\ref{NegVsPtot}). We can see sharpest reduction of ACPE occurring while the coherent evolution become decoherent, but changing $d$ does not affects a lot.(Fig.\ref{NegVsT})  

\section{summary and conclusions}\label{conclusion}
In this paper we derive a general analytic expression for the variance of 1DQW with homogeneous position space, considering the general form of noise over the position space of one dimensional quantum walks.
This expression shows that existence of noise over the position space of a walker, not only does not eliminate $t^2$ term, but also increases the linear term of the variance.

Since, we have derived our expression in asymptotic regime, it seems that increasing the linear coefficient is negligible, but we have shown that the fast decay of time-dependent terms cause these expressions to be correct with a high precision even for the limited number of steps. Specially our simulations show that in some cases for a limited number of steps, noise over the position, can increase the variance significantly.

Presenting an example and the related simulations, we have tested the validity of our expression and showed that noise over the position space does not make transition to a classical system. Especially, entanglement between coin and position (ACPE) of a quantum walker as an another quantum property of system has also been tested for a walker which can tunnel out to its neighbors. We found that, unlike the system with noise on its coin space, ACPE  never vanishes.

Expressions and formulas which we have derived in this paper, not only prove that the environment noise over the position space of homogeneous-position decoherent QW does not change the system to a classical one, but also their generality can be used as a helpful tool for the investigation of the environmental noise effect on the variance of 1DQW.
Furthermore, these relations are calculated generally for any arbitrary initial state, therefore they provide us a precise tool for the effect of the initial states on the variance of the quantum walks.
%
%
%
%

\acknowledgements{ I would sincerely like to thank Maryam Boozarjmehr from the Australian National University, for her accurate reading and editing the  manuscript. }


%

\end{document}